\documentclass{article}

\usepackage{arxiv}

\usepackage[utf8]{inputenc} 
\usepackage[T1]{fontenc}    
\usepackage{hyperref}       
\usepackage{url}            
\usepackage{booktabs}       
\usepackage{amsfonts}       
\usepackage{nicefrac}       
\usepackage{microtype}      
\usepackage{lipsum}
\usepackage{graphicx}
\graphicspath{ {./images/} }
\usepackage{csquotes}
\usepackage{natbib}
\usepackage{tabularx}
\usepackage{rotating}
\usepackage{comment}

\title{Seeing ChatGPT Through Students' Eyes: An Analysis of TikTok Data}

\author{
  Anna-Carolina Haensch \\
  Department of Statistics\\
  LMU Munich\\
  Germany \\
  \texttt{anna-carolina.haensch@stat.uni-muenchen.de} \\
   \And
 Sarah Ball \\
  Department of Statistics\\
  LMU Munich\\
  Germany \\
  \texttt{sarah.ball@stat.uni-muenchen.de} \\
  \And
 Markus Herklotz\\
  Department of Statistics\\
  LMU Munich\\
  Germany \\
  \texttt{markus.herklotz@stat.uni-muenchen.de} \\
   \And
  Frauke Kreuter \\
 Department of Statistics\\
  LMU Munich\\
  Germany \\
  \texttt{frauke.kreuter@stat.uni-muenchen.de} \\
}

\begin{document}
\maketitle
\begin{abstract}
Advanced large language models like ChatGPT have gained considerable attention recently, including among students. However, while the debate on ChatGPT in academia is making waves, more understanding is needed among lecturers and teachers on how students use and perceive ChatGPT. To address this gap, we analyzed the content on ChatGPT available on TikTok in February 2023. TikTok is a rapidly growing social media platform popular among individuals under 30. Specifically, we analyzed the content of the 100 most popular videos in English tagged with \#chatgpt, which collectively garnered over 250 million views. Most of the videos we studied promoted the use of ChatGPT for tasks like writing essays or code. In addition, many videos discussed AI detectors, with a focus on how other tools can help to transform ChatGPT output to fool these detectors. This also mirrors the discussion among educators on how to treat ChatGPT as lecturers and teachers in teaching and grading. What is, however, missing from the analyzed clips on TikTok are videos that discuss ChatGPT 
producing content that is nonsensical or unfaithful to the training data.  

\end{abstract}


\section{Introduction}

ChatGPT represents a significant advancement in the usability of natural language processing technology, which has important implications for a variety of fields such as computer science, law, finance, education, and medicine -- to only name a few \citep{choi2023, dowling2023, kung2023, haque2022}. In particular, ChatGPT's ability to generate texts, such as parts of essays or answers to open-ended questions often indistinguishable from human responses, has made ChatGPT a popular topic in universities and schools \citep{baidoo-anu_education_2023, garcia-penalvo_perception_2023,kasneci_chatgpt_2023,zhang_preparing_2023}.
As more and more industries and fields begin to incorporate AI technologies into their operations, it is becoming increasingly important for students and professionals alike to develop a better understanding of how they can be leveraged to solve complex problems and create new opportunities. As a result, a growing body of literature has focused on exploring the impact of AI language models like ChatGPT on academic teaching and learning (see literature review in \cite{rudolph_chatgpt_2023}). Educators discuss whether the tool should be incorporated or regulated in universities, given the potential of believable AI-generated text to harm academic integrity \citep{cotton_chatting_2023}. 

Expanding on the existing literature, this study focuses on the students' perspective by analyzing social media data. 
Social media has been shown to both shape discourses on educational topics \citep{sam_shaping_2019} and to provide insight into such discourses \citep{curran_sense-making_2017}.
In this study, we investigate TikTok videos to understand how students learn about ChatGPT and its potential applications. TikTok is a popular social media platform that people under 30 widely use. The platform's popularity and ability to disseminate information quickly make it an ideal tool for gathering data about younger persons' perceptions of ChatGPT. 
We will explore what aspects of ChatGPT they find most interesting and how they perceive its potential applications. We gathered data on the 100 most liked TikTok videos with the hashtag \#chatgpt in February 2023, including several metrics like the number of likes, shares, and views. We also developed a classification scheme to get an overview of the content of these videos. Altogether, these videos represent around 250 million views and 22 million likes. In doing so, this study hopes to discover new insights into what students learn about ChatGPT on social media and, even more important, what is not communicated.

This research note is structured into five main sections. We first overview the existing research on ChatGPT in education; and we contextualize the current study within the broader field of research on ChatGPT and education. The next section describes the research design and data collection process. The results section presents the most popular themes and topics discussed in the collected TikTok videos and several count metrics. Finally, the discussion section will explore how the study's results contribute to our understanding of students' engagement with ChatGPT and the implications of these findings for educational practices.

\section{ChatGPT and higher education}
While ChatGPT has made an impact in many areas, it seems to have left an especially big mark in secondary and tertiary education. We traced the conversations and discussions on the side of educators through the numerous publications that have emerged since the release of ChatGPT in November 2022.

One of the main discussion points in the literature are ChatGPT's capabilities and weaknesses regarding exams, essays, and other forms of examination. As an example, in the United States Medical Licensing Examination ChatGPT performed at a level of a third-year medical student \citep{gilson_how_2023}. \citet{jalil_chatgpt_2023} confronted ChatGPT with questions from a well-known text book on software testing. The tool answered either completely or partially correctly in 44\% of the cases. Another study found that ChatGPT performs similarly to web-based search when used as a writing assistant for student essays, which is why the study authors saw no need for alarmed reactions among educators \citep{basic_better_2023}.

Several studies also examined ChatGPT's capabilities in creating scientific texts, and their differences to those written by humans. \citet{kutela_chatgpts_2023} were able to distinguish human-generated from ChatGPT-generated manuscripts by applying different supervised text mining approaches. While \citet{gao_comparing_2022} noted that ChatGPT can create believable scientific abstracts, they were also able to identify them as AI-generated in the majority of cases. This was true for AI- and plagiarism-detectors, but also for (less accurate) human reviewers. Other studies which trained models to differentiate between human- and ChatGPT-generated text were also successful, although with varying accuracy \citep{mitrovic_chatgpt_2023,shijaku_chatgpt_2023}.

Now that ChatGPT and similar large language models are widely available as tools, the discussion on academic integrity is also very active \citep{cotton_chatting_2023,khalil_will_2023,rudolph_chatgpt_2023}. One important aspect of the discussion is the perceived eloquence of text generated with ChatGPT combined with the tendency of ChatGPT to \enquote{hallucinate}, that is, to invent facts that are not correct. To effectively utilize AI tools, users must engage in a critical assessment of the text they produce. This assessment, in turn, will become a significant task for students  \citep{bishop_computer_2023}. There are already guides for teachers and learners on how to use ChatGPT \citep{atlas_chatgpt_nodate}, and \citet{willems_chatgpt_2023} argues that it is the educators' responsibility to instruct students to use it responsibly, just like any other tool. \citet{kasneci_chatgpt_2023} are convinced that ChatGPT itself could provide the necessary training to accustom students to the potential biases and risks of AI applications. 

The conclusion shared by several authors is that ChatGPT is here to stay, and simply banning it does not seem very promising \citep{garcia-penalvo_perception_2023,baidoo-anu_education_2023}. Rather, it is a chance to reflect on current teaching practices under these new developments and to focus on the students' perspective. Instead of just ignoring the capabilities and possible applications of the tool, educators should give students tasks that they believe are worth doing \citep{zhang_preparing_2023}.

However, to capture the discussion among students and identify differences to the discussion among educators, we have to look beyond academic publications for now. Simply surveying students could prove difficult, as the necessary data preparation for extensive surveys often takes several months. In addition, social desirability bias could be a problem, as the discourse on plagiarism and cheating may lead students to distance themselves from using ChatGPT when responding to surveys on ChatGPT. Therefore, we will look at alternative data sources, that is, TikTok data, in our case. As the most popular social media platform for younger generations, current topics are discussed from and with their perspective on this Social Media platform. TikTok is increasingly discussed in terms of educational content \citep{rahimullah_assessing_2022}, and through the sheer representation of the relevant age groups \citep{statista_tiktok_brand_report_2021}, the platform shapes perceptions of ChatGPT for these groups. We will discuss this data source in more detail in the next section.

\section{Data and Methods incl previous research with TikTok data}

\paragraph{TikTok as a platform}

TikTok is a social media platform that allows users to create and share short video clips. TikTok was first launched in China in 2016 under the name Douyin. Outside of China, TikTok was originally a separate platform known as musical.ly, which was then bought by Douyin and rebranded outside of China as TikTok in 2018. It has become one of the world's most popular social media platforms, with a billion users every month. TikTok users can create and watch short videos ranging from 15 seconds to 10 minutes. They can also use various creative tools such as filters, music, and templates to edit their videos. Its algorithm-based content delivery system sets TikTok apart from many other platforms. TikTok shows less content from a user's personal network than other social media platforms, and strongly tailors content to a user's expressed preferences and past engagement with content \citep{anderson2020getting}. The platform is especially popular among younger audiences aged 13-29. According to data from the United States, 32.5\% of TikTok users are between the ages of 10-19, while 29.5\% are aged 20-29 \citep{clement2020ustiktok}.

\paragraph{Research with TikTok data}
Research studies using TikTok data usually analyze the content available on TikTok. They often aim to understand better what younger generations consume content-wise, especially regarding sensitive topics such as mental health in a pandemic. The studies included in a recent systematic review on TikTok studies by \citet{mccashin2023using} covered many issues, including COVID-19, dermatology, eating disorders, cancer, tics, radiology, sexual health, DNA, and public health promotion. Most studies originated from the United States; the rest were from China, Ireland, Australia, and Canada. TikTok has become a leading social media platform for American teenagers aged 18 to 19, with 67\% of them reporting using it \citep{statista_tiktok_2021}. Obtaining the share of college-aged persons that use TikTok is harder, but it is estimated that 56\% of persons aged 20-29 use the platform \citep{statista_tiktok_2021}. TikTok also has a higher share of 18 - 29-year-old users than other social networks \citep{statista_tiktok_brand_report_2021}. Therefore, if one is mainly interested in studying young people's behaviors, attitudes, or preferences, TikTok can be a potential data source. Still, it is essential to remember that the sample may not represent the overall population or specific subgroups within it. However, a recent study shows no bigger differences between users and non-users of TikTok regarding variables such as income, education, household composition, and the type of community they live in \cite {statista_tiktok_brand_report_2021}. 

\paragraph{Sample}

Regarding the data collection and analysis, we use similar strategies as other studies using TikTok data \citep{Basch2021, Fiallos2021, Fowler2022, McCashin2022, mccashin2023using}. The 100 most popular videos (measured by likes) were selected from the hashtag \#chatgpt, which had a total of 1.3 billion views at the time of the study. The data were collected in one day (7th February 2023). 
To gain the top 100 videos on ChatGPT in English, we had to screen 173 videos and excluded in total 73 videos that were not in English or not on the topic of ChatGPT.

\paragraph{Measurement}

We collected the following variables for all videos: cumulative views and the number of likes and shares, as well as the content creators' ID. Then, to categorize the content, we iteratively developed a coding system. Two authors screened around 20 videos each, before agreeing on a classification scheme, and then coded the top 100 videos on ChatGPT in English manually. The authors agreed on the following classification scheme for the content presented in the TikTok videos, which can be seen in Table \ref{category_table}.

The classification scheme is organized into several major categories. 
The first major category is \textit{Promotional}, which includes videos explaining how to use ChatGPT in general and accomplish specific tasks. It also encompasses tutorials on how to use ChatGPT more efficiently (Prompt engineering) or how to use ChatGPT without getting caught by AI detectors. The category also includes subcategories for writing code in different programming languages (Python, Excel, C++), writing essays or other text types, getting summaries, and generating research ideas.

The second major category, called \textit{Critical}, includes videos criticizing ChatGPT. This category contains subcategories such as failures to provide correct answers, server breakdowns, content filters deemed too restrictive, and videos stressing that ChatGPT output can be detected by AI detection software, thereby posing a threat to students who use ChatGPT in their work. Additionally, while these videos mention the listed problems, they do not promote other tools or remedies to circumvent them.

The other categories are \textit{Business with ChatGPT}, that is, videos that examine ways in which AI language models can be used for business purposes; \textit{The future of society}, which describes videos that explore the potential impact of AI language models on society; \textit{AI tools list}, a video format in which different AI language models are contrasted in list form; and \textit{Entertainment}, that is, videos in which ChatGPT is used as the topic of a humorous sketch. We only included subcategories for the first two categories since the other categories did not have as many observations, and they are less interesting for our research purpose.

Additionally, we classified whether the videos show content that is directly related to academic pursuit or not. For instance, videos that explain how to use ChatGPT for assignments, exams, or papers in education would be classified as related to academic pursuit.

 \begin{table}[!h]
 \centering
 \caption{Categories and subcategories for the top 100 TikTok videos with the hashtag \#ChatGPT.  \label{category_table}}
 \begin{tabular}{lll}
 \toprule
 \textbf{Category} &  \textbf{Subcategory} & \textbf{Count} \\
 \midrule
Promotional  &  Write essays & 7\\
& Write other texts (letters, poems, recipes) & 17\\
& Write code & \\
&   \quad Python & 3\\
&   \quad Excel & 1\\
&   \quad C++ & 1\\
& \quad others (HTML, CSS, JavaScript) & 3 \\
& Answer questions & 6\\
& Get research ideas & 1\\
& Prompt engineering & 4 \\
& Take meeting notes from recording & 1 \\
& Fool AI detectors  \\
&   \quad Quillbot  & 6\\
&   \quad Chatsonic  & 1\\
&   \quad Tutorly.Ai & 2 \\
 \midrule
Critical &    AI detectors as threat  \\
&   \quad AI detectors in general  & 2\\
&   \quad GPTZero  & 4\\
&Failures to provide correct answers & 1 \\
& Server breakdowns & 1 \\
& Filters too restrictive & 2 \\
& Proposing alternative to ChatGPT: Caktus & 2 \\
 \midrule
Business with ChatGPT & &2  \\
 \midrule
The future of society  && 10  \\
 \midrule
AI tools list & & 5\\
 \midrule
Entertainment & & 18 \\
 \bottomrule
 \end{tabular}
 \end{table}

 One coder (SB) coded all videos by hand, and a second coder (ACH) coded a random sample of 54 observations to assess inter-rater reliability: Cohen's kappa is 0.896.

\clearpage

\section{Results}
Moving on to the results of our analysis, the following section first describes the video content, which is followed by the analysis of count numbers related to likes, shares, views, and the video creators.

\textbf{Video content} 
Of the 100 selected TikTok videos, 53 are promotional, meaning they present ChatGPT and its applications in a positive light. Given the short nature of the videos, the majority of the promotional videos quickly showcase the capabilities of ChatGPT like answering questions (6), essay writing (7 videos), writing code (8 videos), or other types of text such as letters, poems or recipes (17 videos). However, some videos engage with ChatGPT in a more detailed way by giving more advanced tutorials on how to use the tool. For instance, there is a video about how to obtain new research ideas or how to transform a recording into written meeting notes. Other videos show how to best design the prompt to get the desired output (4 videos). Furthermore, there is a focus on how to fool AI detectors with tools that transform the ChatGPT output, like Quillbot (6 videos), Chatsonic (1 video), and Tutorly.Ai (2 videos). 

Compared to the category of promotional videos, far fewer clips are critical of ChatGPT (12 videos). The majority in this class forms the subcategory “AI detectors as threat” (6 videos); with the student-built software \textit{GPTZero} being the main topic in within this subcategory (4 videos). The other videos are about a failure to provide correct answers (1 video), ChatGPT server breakdowns (1 video), and criticism that the content filters imposed by OpenAI are too restrictive (2 videos). Furthermore, two videos suggest using another text generation tool called Cactus and claim it is better at writing essays than ChatGPT.

Another important category is “entertainment” (18 videos), under which all videos are summarized that mainly provide humorous content about the usage of ChatGPT. Furthermore, many clips fall under the category “the future of society” (10 videos). In those clips, people predict how ChatGPT might change society, including education. Lastly, five videos show lists of AI tools that can be used to do certain tasks, and two videos present how one can use ChatGPT to do business.

Of the 100 videos, 40 were directly related to academic pursuit. The share of education-related videos among the promotional videos is 37.5\%, while it is 75\% for the critical videos. Hence, critical videos are related to academic assignments far more often than promotional videos.

\begin{figure}[h]
\includegraphics[width=\textwidth]{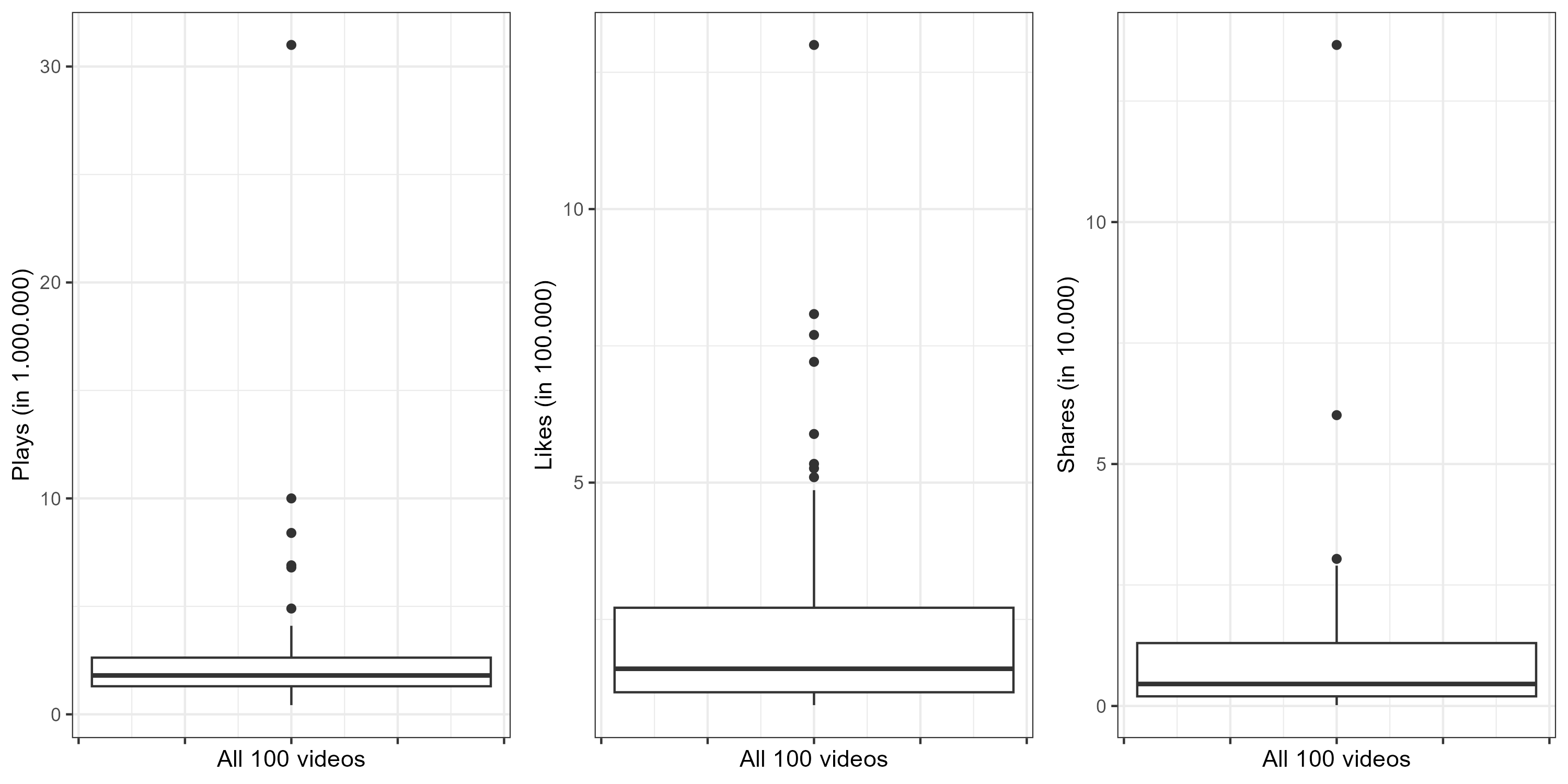}
\caption{Play, like, and share count for all the selected videos. \label{box_all}}
\end{figure}

\textbf{Likes, Shares, and Views} The videos in our dataset cover a total of 22,760,300 likes, 922,041 shares, and 248,522,100 views. The average video was liked 227,603 times ($max = 1,300,000 ; min = 93,300$), shared 9,220 times ($max = 136,600 ; min = 178$) and played 2,485,221 times ($max = 31,000,000 ; min = 426,100$), see also Figure \ref{box_all}. 

\begin{figure}[h]
\includegraphics[width=0.85\textwidth]{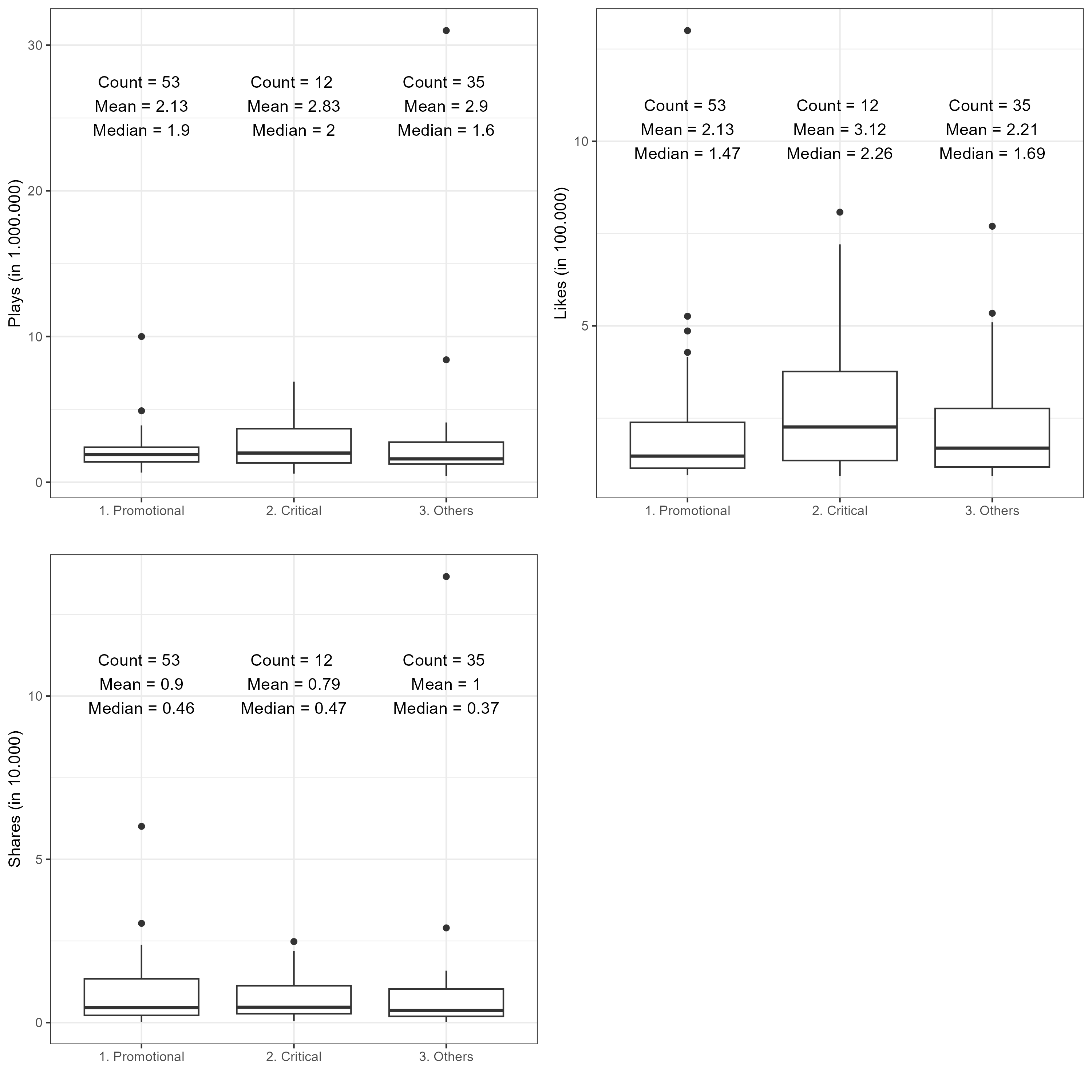}
\caption{Play, like and share count for the selected videos, grouped by category. \label{box_prom_critical}}

\end{figure}

When looking at promotional and critical videos separately, promotional videos were liked 11,293,300 times, shared 477,670 times, and viewed 113,072,700 times, while critical videos were liked 3,749,000 times, shared 95,224 times, and viewed 33,946,500 times, see also Figure \ref{box_prom_critical}.

Looking at Table \ref{tab:videos}, the \textit{average} critical TikTok video received more views and had higher levels of approval through likes than the average promotional video; the frequency of sharing was, however, comparatively lower.\footnote{A Welch two sample t-test revealed no significant differences concerning the means of those counts between the promotional and critical videos.} Nevertheless, in total, much more people have seen, liked and shared promotional videos than critical videos about ChatGPT.

\begin{table}[htbp]
\centering
\caption{Promotional and critical videos statistics \label{tab:videos}}
\resizebox{\textwidth}{!}{%
\begin{tabular}{|l|c|c|c|}
\hline
\multicolumn{1}{|c|}{\textbf{Category}} & \textbf{Likes} & \textbf{Shares} & \textbf{Views} \\ \hline
Average Promotional & 213,081  & 9,013  & 2,133,447  \\ 
 & ($\max = 1,300,000; \min = 95,700$) & ($\max = 60,100; \min = 178$) & ($\max = 10,000,000; \min = 659,000$) \\ \hline

Average Critical & 312,417  & 7,935 & 2,828,875  \\ 
& ($\max= 808,100; \min = 93,700$) & ($\max = 24,800; \min = 526$) &($\max = 6,900,000; \min = 587,900$) \\ \hline \hline
Total promotional & 11,293,300 & 477,670 & 113,072,700 \\ \hline
Total critical & 3,749,000 & 95,224 & 33,946,500 \\ \hline
\end{tabular}}
\end{table}

\textbf{Video creators}
In our selection of videos, only eight creators have posted more than one video. Among the eight most active video posters, the highest number of released videos is seven, followed by the second-highest number of four. Hence, there is some but no extreme concentration of specific content creators in the top 100 videos.

\clearpage

\section{Discussion}

This study explored what students can learn about ChatGPT and its potential applications through the social media platform TikTok. To gain insight into the most popular themes and topics discussed in these videos, we analyzed the 100 most liked TikTok videos with the hashtag \#chatgpt in English. The results indicate that ChatGPT's ability to generate human-like texts is intriguing for many platform users.
A lot of the videos have a positive outlook on ChatGPT and focus on actual applications, such as writing essays and other texts, providing code, and answering questions.

The academic debate measured through the published literature is centered around concerns and challenges regarding ChatGPT. Discussion points include ChatGPT being used as a tool for cheating and plagiarism, for providing users with wrong information, and students not achieving their learning objectives when using ChatGPT as they do not have to understand and complete tasks themselves. The analyzed content on TikTok (even if one focuses only on videos related to academic assignments) has remarkably different ranges. The vast majority of the videos promoted the usage of ChatGPT without any hesitation and those videos also had higher like, share, and view counts compared to the critical videos. In total, only one of the 100 TikTok videos highlights failures of ChatGPT to provide correct answers, even though 13 videos present ChatGPT as a tool for essay writing or to answer questions. Furthermore, although a considerable number of videos mention that ChatGPT-produced text has the disadvantage of being detectable by AI software, those videos mainly focus on how to circumvent such a potential drawback of ChatGPT.

The lack of discussion around limitations related to the use of ChatGPT in the analyzed TikTok videos is a concern that requires further exploration. 
 From what we have seen, the risks of producing eloquent but incorrect responses and similar limitations of ChatGPT like inherent biases of the produced output are currently very underrepresented on the most popular social media platform among younger people. We therefore want to stress the importance of promoting responsible AI usage in educational settings, including discussions about ethical considerations and limitations.
 Contrary to simply banning the tool from schools and campuses, including such issues in curricula allows setting standards and achieve academic integrity while staying up-to-date with current developments. As shown by the videos on TikTok, students will probably get to know ChatGPT one way or the other, but educators now have the chance to introduce the tool responsibly.

In summary, this study provides insights into what students can learn about ChatGPT on one of the most popular social media platforms. Even more importantly, we also identified potential gaps in their knowledge. We furthermore showed how researchers can leverage the popularity and accessibility of TikTok to explore what younger generations discuss and think about different trending topics. It is, however, still important to consider potential measurement and sampling errors of our study. Regarding sampling, using TikTok data to assess students' perceptions of ChatGPT may not accurately represent the broader population of students, as TikTok users may not represent the larger student population. However, studies examining the TikTok user base show no concerning differences, as discussed in the section on data collection. Looking at the measurement side, we measured the information readily available to students through the content of the top 100 English TikTok videos with the hashtag \#chatgpt. This approach ignores all the other videos related to ChatGPT on the TikTok platform, and also all alternative sources of information that students might access to learn about and engage with ChatGPT. Despite these limitations, our main results are essential for educators who want to encourage and teach their students how to use ChatGPT responsibly. These findings include the neglect in the most trending videos to address ChatGPT's tendency to hallucinate, as well as their strong focus on essay writing with ChatGPT and fooling AI detectors.  

\newpage

\bibliographystyle{apalike}

\end{document}